\documentstyle[aps,12pt,preprint]{revtex}
\tightenlines

\begin{document}
\title{{\bf Pauli's Term and Fractional Spin}}
\author{F. A. S. Nobre\thanks{%
Permanent address: Universidade Regional do Cariri-URCA, Crato-Ce, Brazil},\
C. A. S. Almeida}
\address{{\normalsize {\it Universidade Federal do Cear\'{a}}}\thinspace \\
{\normalsize {\it Physics Department}}\\
{\normalsize {\it C.P. 6030, 60470-455 Fortaleza-Ce, Brazil}}\thanks{%
Electronic address: augusto@cbpf.br, carlos@fisica.ufc.br}}
\maketitle

\begin{abstract}
In this work we consider an Abelian Chern-Simons-Higgs model
coupled non-minimally to matter fields. This coupling is
implemented by means of a Pauli-type coupling. We show that the
Pauli term is sufficient to gives rise to fractional spin.
\end{abstract}

\vspace{0.2cm} PACS numbers: 11.15.-q, 11.10.Ef, 11.10.Kk, 11.30.-j \vspace*%
{-0.5pt}

\vspace{1.0cm}

The fractional statistics\cite{wilczek} with its theoretical and
applicable consequences plays an interesting interplay role
between quantum field theory and condensed matter physics.
Previous speculations\cite {laughlin1} that the fractional quantum
Hall effect could be explained by quasiparticles (anyons) obeying
fractional statistics were confirmed both numerically and
analytically\cite{hall1}. So far not so successfully, anyons
are also studied in the context of high $T_{c\text{ }}$ superconductivity%
\cite{laughlin}.

As it is known, the presence of Chern-Simons terms in (2+1)
dimensional gauge theories induce fractional
statistics\cite{semenoff,dunne}. However, besides the statistical
interaction, anyons may interact electro-magnetically\cite {hong}.
To describe the fractional quantum Hall effect, a bound state of
anyon and a magnetic flux tube is required\cite{jain}, which means
that the interaction between the electric charge and the magnetic
moment of the anyon must be considered.

Chern-Simons-like theories involving an anomalous magnetic moment
have attracted much attention in the literature in recent years.
Stern \cite {stern} was the first, as far as we know, to suggest a
nonminimal term in the context of the Maxwell-Chern-Simons
electrodynamics with the intention of mimicking an anyonic
behavior without a pure Chern-Simons limit. This term can be
interpreted as a tree level Pauli-type coupling, {\it i. e.}, an
anomalous magnetic moment. It is a specific feature of (2+1)
dimensions that the Pauli coupling exists, not only for spinning
particles, but also for scalar ones \cite{torres}. As a
fundamental result Stern showed that, for a particular value of
the nonminimal coupling constant, the field equations of his model
coincide with the field equations of a {\it pure} Chern-Simons
theory minimally coupled.

In this Letter, we consider an Abelian Chern-Simons-Higgs theory
where the complex scalar fields couples directly to the
electromagnetic field strength (Pauli-type coupling). We quantize
canonically the theory using the Dirac brackets method and compute
the angular momentum operator with the contribution of the Pauli
term. Finally, we show that this term gives rise to fractional
spin even in the absence of Chern-Simons term.

The Lagrangian of the model under investigation is

\begin{equation}
L=\left| \nabla _\mu \phi \right| ^2+\frac \kappa 2\varepsilon ^{\mu \nu
\lambda }A_\mu \partial _\nu A_\lambda -A_\mu \partial ^\mu b+\frac \alpha 2%
b^2  \label{eq.1}
\end{equation}
where $\nabla _\mu \phi \equiv (\partial _\mu -ieA_\mu -i\frac g4\varepsilon
_{\mu \lambda \sigma }F^{\lambda \sigma })\phi $. Note that this covariant
derivative includes both the usual minimal coupling and the contribution due
to Pauli's term. Here $A_\mu $ is the gauge field and the Levi-Civita symbol $%
\varepsilon _{\mu \nu \lambda }$ is fixed by $\varepsilon _{012}=1$ and $%
g_{\mu \nu }=diag(1,-1,-1)$. The multiplier field $b$ has been introduced to
implement the covariant gauge-fixing condition.

Before quantizing the theory, we analyze the above Lagrangian in
terms of Hamiltonian methods. Here we follow the approach used by
Shin {\it et al.} \cite{kim}. We carry out the constraint analysis
of this model, in order to obtain a consistent formulation of the
theory.

The canonical momenta of the Lagrangian (\ref{eq.1}), which can be easily
seen by considering its temporal and spatial components separately, are
given by


\begin{equation}
\pi _0=0\text{ },  \label{eq.2a}
\end{equation}

\begin{equation}
\pi _b=-A_0\text{ ,}  \label{eq.2b}
\end{equation}

\begin{equation}
\pi ^j=-\frac \kappa 2\varepsilon ^{ij}A_i-i\frac g2\varepsilon ^{ij}\left[
\phi ^{*}(D_i\phi )-\phi (D_i\phi )^{*}\right] -\frac{g^2}2\partial
^jA_0\left| \phi \right| ^2+\frac{g^2}4(\partial _0A_0)\left| \phi \right| ^2%
\text{ ,}  \label{eq.2c}
\end{equation}

\begin{equation}
\pi =(\partial _0\phi ^{*})+ieA_0\phi ^{*}+i\frac g4\phi ^{*}\varepsilon
^{ij}F_{ij}\text{ ,}  \label{eq.2d}
\end{equation}

\begin{equation}
\pi ^{*}=(\partial _0\phi )-ieA_0\phi -i\frac g4\phi \varepsilon ^{ij}F_{ij}%
\text{ ,}  \label{eq.2e}
\end{equation}
where $\pi _0$, $\pi ^j,$ $\pi _{b\text{ }},$ $\pi $ and $\pi ^{*}$ are the
canonical momenta conjugate to $A_0,$ $A_j$, $b,$ $\phi $ and $\phi ^{*}$
respectively. Also we have used $\varepsilon _{ij}=\varepsilon _{0ij}$ , $%
D_i=$ $\partial _i-ieA_{i\text{ }}$and $i,j=1,2$ .

The canonical momenta (\ref{eq.2a}) and (\ref{eq.2b}) do not involve
explicit time dependence and hence are primary constraints. Performing the
Legendre transformation, the canonical Hamiltonian can be written as

\begin{eqnarray}
H_c &=&\pi ^{*}\pi +\left| D\phi \right| ^2+ieA_0(\pi \phi -\pi ^{*}\phi
^{*})+\kappa \varepsilon ^{ij}A_0\partial _iA_j+A_i\partial ^ib-\frac \alpha %
2b^2  \nonumber \\
&&-i\frac g2\varepsilon ^{ij}\partial _jA_0\left[ \phi ^{*}(D_i\phi )-\phi
(D_i\phi )^{*}\right] -\frac{g^2}4\partial _iA_0\partial ^iA_0\left| \phi
\right| ^2  \nonumber \\
&&-\frac g4\varepsilon ^{ij}F_{ij}\left[ \phi ^{*}(D_0\phi )-\phi (D_0\phi
)^{*}\right] -\frac{g^2}8F^{ij}F_{ij}\left| \phi \right| ^2.  \label{eq.3}
\end{eqnarray}

Now, in order to implement the primary constraints in the theory, we
construct the primary Hamiltonian as

\begin{equation}
H_p=H_c+\lambda _0\pi +\lambda _1(\pi _b+A_0)\text{ ,}  \label{eq.4}
\end{equation}
where $\lambda _0$ and $\lambda _1$ are Lagrange multiplier fields.
Conserving in time the primary constraints yields the secondary constraints

\begin{equation}
\psi _1=\pi _0\approx 0\text{ ,}  \label{eq.5a}
\end{equation}

\begin{equation}
\psi _2=\pi _b+A_0\approx 0\text{ ,}  \label{eq.5b}
\end{equation}
which are also conserved in time and where the symbol $\approx $ indicates
weak equality, {\it i. e.}, the constraints can be identically set equal to
zero only after computing the relevant Poisson brackets. Thus there is no
more constraint and the above equations are the set of fully second-class
constraints. On the other hand, there is no first-class conditions and so,
no gauge conditions to be determined in theory. This is an effect of the
gauge fixing condition imposed previously. As it is known, the lack of
physical significance allows that the second-class constraints can be
eliminated by means of Dirac brackets (DB's).

Following the standard Dirac brackets formalism and quantizing the
system, we obtain the following set of non-vanishing equal-time
commutators:

\begin{equation}
\left[ A_0(x),b(y)\right] =i\delta ^2(x-y)  \label{eq.6a}
\end{equation}

\begin{equation}
\left[ A_i(x),\pi _j(y)\right] =i\delta _{ij}\delta ^2(x-y)  \label{eq.6b}
\end{equation}

\begin{equation}
\left[ \phi (x),\pi (y)\right] =\left[ \phi ^{*}(x),\pi ^{*}(y)\right]
=i\delta ^2(x-y)  \label{eq.6c}
\end{equation}

After achieving the quantization we proceed to construct the angular
momentum operator and compute the angular momentum of the matter field $\phi
$.

The symmetric energy-momentum tensor can be obtained by coupling the fields
to gravity and then varying the action with respect to $g^{\mu \nu }$:

\begin{eqnarray}
T_{\mu \nu } &=&\frac 2{\sqrt{-g}}\frac{\delta S}{\delta g^{\mu \nu }}
=(\nabla _\mu \phi )^{*}(\nabla _\nu \phi )+(\nabla _\nu \phi )^{*}(\nabla
_\mu \phi )-A_\mu \partial _\nu b-A_\nu \partial _\mu b  \nonumber \\
&&-g_{\mu \nu }(\left| \nabla _\alpha \phi \right| ^2-A_\alpha \partial
^\alpha b)\text{ .}  \label{eq.7}
\end{eqnarray}

The angular momentum operator in (2+1) dimensions is given by

\[
L=\int d^2x\varepsilon ^{ij}x_iT_{0j}\text{ .}
\]
Hence

\begin{eqnarray}
L &=&\int d^2x\varepsilon ^{ij}x_i\{(\pi \partial _j\phi +\pi ^{*}\partial
_j\phi ^{*})-ieA_jJ_0-i\frac g2\varepsilon _{jl}F^{l0}(\pi \phi -\pi
^{*}\phi ^{*})-A_0\partial _jb  \nonumber \\
&&+A_j\partial _0b-i\frac g2A_j\varepsilon ^{kl}\partial _k[\phi
^{*}(D_l\phi )-\phi (D_l\phi )^{*}]+i\frac{g^2}2A_j\partial _k(\left| \phi
\right| ^2F^{0k})\}\text{ ,}  \label{eq.8}
\end{eqnarray}
where

\begin{equation}
J_0=i\{\pi \phi -\pi ^{*}\phi ^{*}-\frac g{2e}\varepsilon ^{ij}\partial
_i[\phi ^{*}(D_j\phi )-\phi (D_j\phi )^{*}]+i\frac{g^2}{2e}\partial
_i(\left| \phi \right| ^2F^{0i})\}  \label{eq.9}
\end{equation}
is the temporal component of the conserved matter current.

The key point here is that Gauss' law is no more a constraint,
while $J_0$ and $T_{\mu \nu }$ contain derivatives of $A_\mu $ .
Note that, due to its topological character, the Chern-Simons term
does not contribute to the energy-momentum tensor. These aspects
are attributed to the non-linearity introduced by Pauli's term.

The rotational property of the $\phi $ field is obtained by computing the
commutator $[L,\phi (y)]$. Using equations (\ref{eq.6a}-\ref{eq.6c}) and (%
\ref{eq.8}), it is easy to see that

\begin{equation}
\lbrack L,\phi (y)]=\varepsilon ^{ij}y_i\partial _j\phi -[e\int
d^2x\varepsilon ^{ij}x_iA_jJ_0,\phi ]+i\frac g2\varepsilon ^{ij}\varepsilon
_{jk}y_iF^{k0}\phi \text{ .}  \label{eq.10}
\end{equation}

This commutator can be rewritten by means of the electromagnetic charge
operator

\[
Q=\int d^2xJ_0(x)
\]
and becomes

\begin{equation}
\lbrack L,\phi (y)]=\varepsilon ^{ij}y_i\partial _j\phi -\frac{e^2}{4\pi
\kappa }[Q^2,\phi (y)]+i\frac g2\varepsilon ^{ij}\varepsilon
_{jk}y_iF^{k0}\phi \text{ }  \label{eq.11}
\end{equation}
or, in more familiar notation

\begin{equation}
\lbrack L,\phi (y)]=i({\bf y}\times {\bf \nabla })\phi (y)-\frac{e^2}{2\pi
\kappa }Q\phi (y)+i\frac g2{\bf y\cdot E}\phi (y)\text{ .}  \label{eq.12}
\end{equation}

The first term in the right hand side of eq. (\ref{eq.12})
represents the intrinsic spin and the second is the so-called
rotational anomaly, which is responsible for the fractional spin.
The term that involves the electric field is the central point of
this work. Unlike the Chern-Simons term (whose contribution is
related with magnetic field), the Pauli term induces an anomalous
contribution for the spin of the system, which depends on electric
field. We stress that, here the nonminimal coupling constant is a
free parameter.

It is worth mentioning that all the procedure above can be carried
out even if there is no Chern-Simons term in the Lagrangian
(\ref{eq.1}). In this case the anomalous contribution to spin
would just come from the Pauli term.

Now we will discuss the above result in connection with theories in the
broken-symmetry phase. Boyanovsky \cite{boy} has found that the low-lying
excitations of a U(1) Chern-Simons theory in interaction with a complex
scalar field in a broken symmetry state are massive bosons with {\it %
canonical statistics}. He explained his result as due to the
screening of long-range forces in a broken symmetry phase. In this
phase localized charge distributions cannot be supported, which is
supposed to be essential for fractional spin. On the other hand,
if we consider a non-minimally coupled Abelian-Higgs model, the
long-distance damping effect by the ''photon'' mass $\kappa $ no
longer exists. This is an indication that Pauli's term, which
induces an anomalous\- spin, can be relevant for the study of
broken symmetry states (superfluid) in the context of effective
theories in condensed matter.

We conclude with two comments:

a) In nonrelativistic limit, Carrington and Kunstatter \cite{carr} have
shown that anomalous magnetic moment interactions gives rise to both the
Aharonov-Bohm and Aharonov-Casher effects. They have speculated possible
anomalous statistics without the CS term. As a matter of fact, we believe
that this (in a relativistic theory) was proved here.

b) The Abelian Chern-Simons term can be generated by means of a spontaneous
symmetry breaking of a nonminimal theory \cite{latinski,carr}. This
connection between Chern-Simons and Pauli-type coupling was pointed out by
Stern \cite{stern}. So the Pauli term at tree-level (with the nonminimal
coupling constant $g$ as a free parameter) can constitute an effective
theory which bring us information about physical models in broken symmetry
phase.

\end{document}